\documentclass[showpacs,preprintnumbers,amsmath,amssymb]{revtex4}
\begin{document}

\textbf{The Proof that the Standard Transformations of }$\mathbf{E}$ \textbf{%
and} $\mathbf{B}$\textbf{\ and the Maxwell}

\textbf{Equations with }$\mathbf{E}$ \textbf{and} $\mathbf{B}$\ \textbf{are
not Relativistically Correct}\bigskip

\qquad Tomislav Ivezi\'{c}

\qquad \textit{Ru%
\mbox
 {\it{d}\hspace{-.15em}\rule[1.25ex]{.2em}{.04ex}\hspace{-.05em}}er Bo\v
{s}kovi\'{c} Institute, P.O.B. 180, 10002 Zagreb, Croatia}

\textit{\qquad ivezic@irb.hr\bigskip }

In this paper it is exactly proved that the standard transformations of the
three-dimensional (3D) vectors of the electric and magnetic fields $\mathbf{E%
}$ and $\mathbf{B}$ \emph{are not} relativistically correct transformations.
Thence the 3D vectors $\mathbf{E}$ and $\mathbf{B}$ \emph{are not}
well-defined quantities in the 4D spacetime and, contrary to the general
belief, the usual Maxwell equations with the 3D $\mathbf{E}$ and $\mathbf{B}$
\emph{are not} in agreement with the special relativity. The 4-vectors $%
E^{a} $ and $B^{a},$ as well-defined 4D quantities, are introduced instead
of ill-defined 3D $\mathbf{E} $ and $\mathbf{B.}$ The proof is given in the
tensor and the Clifford algebra formalisms. \bigskip

\noindent PACS numbers: 03.30.+p \bigskip

It is generally accepted by physics community that there is an agreement
between the classical electromagnetism and the special relativity (SR). The
standard transformations of the three-dimensional (3D) vectors of the
electric and magnetic fields, $\mathbf{E}$ and $\mathbf{B}$ respectively,
are considered to be the Lorentz transformations (LT) of these vectors, see,
e.g., [1], [2] Sec. 11.10, or [3] par.24. The usual Maxwell equations (ME)
with the three-vectors (3-vectors) $\mathbf{E}$ and $\mathbf{B}$ are assumed
to be physically equivalent to the field equations (FE) expressed in terms
of the electromagnetic field tensor $F^{ab}.$ In this paper it will be
exactly proved that the above mentioned standard transformations of $\mathbf{%
E}$ and $\mathbf{B}$ \emph{are not} relativistically correct transformations
in the 4D spacetime and consequently that the usual ME with $\mathbf{E}$ and
$\mathbf{B}$ and the FE with $F^{ab}$ \emph{are not} physically equivalent.
The whole consideration will be mainly presented in the tensor formalism
(TF), since it is better known, and only briefly in the Clifford algebra
formalism (CAF). It will be shown that in the 4D spacetime the well-defined
4D quantities, the 4-vectors of the electric and magnetic fields $E^{a}$ and
$B^{a}$ in the TF (as in [4,5]), or, e.g., the 1-vectors $E$ and $B$ in the
CAF (as in [6]), have to be introduced instead of ill-defined 3-vectors $%
\mathbf{E}$ and $\mathbf{B.}$

Let us start with some general definitions. The electromagnetic field tensor
$F^{ab}$ is defined without reference frames, i.e., it is an abstract
tensor, a geometric quantity; Latin indices a,b,c, are to be read according
to the abstract index notation, as in [7] and [4,5]. When some reference
frame (a physical object) is introduced and the system of coordinates (a
mathematical object) is adopted in it, then $F^{ab}$ can be written as a
coordinate-based-geometric quantity (CBGQ) containing components and a
basis. The system of coordinates with the Einstein synchronization of clocks
and Cartesian spatial coordinates (it will be called the Einstein system of
coordinates (ESC)) is almost always chosen in the usual treatments. (In my
approach to SR that uses 4D quantities defined without reference frames,
[4,5] and [8] in the TF, and [6] in the CAF, any permissible system of
coordinates, not necessary the ESC, can be used on an equal footing.) When $%
F^{ab}$ is written as a CBGQ (with the ESC) it becomes $F^{ab}=F^{\mu \nu
}\gamma _{\mu }\otimes \gamma _{\nu },$ where Greek indices $\mu ,\nu $ in $%
F^{\mu \nu }$ run from 0 to 3 and they denote the components of the
geometric object $F^{ab}$ in some system of coordinates, here the ESC, $%
\gamma _{\mu }$ are the basis 4-vectors (not the components) forming the
standard basis $\left\{ \gamma _{\mu }\right\} $ and $^{\prime }\otimes
^{\prime }$ denotes the tensor product of the basis 4-vectors. In the TF I
shall often denote the unit 4-vector in the time direction $\gamma _{0}$ as $%
t^{b}$ as well. Then in some reference frame (with the ESC, i.e., with the
standard basis $\left\{ \gamma _{\mu }\right\} $) $t^{b}$ can be also
written as a CBGQ, $t^{b}=t^{\mu }\gamma _{\mu }$, where $t^{\mu }$ is a set
of components of the unit 4-vector in the time direction ($t^{\mu }=\left(
1,0,0,0\right) $). Almost always in the standard covariant approaches to SR
one considers only the components of the geometric quantities taken in the
ESC, and thus not the whole tensor. However the components are coordinate
quantities and they do not contain the whole information about the physical
quantity.

In the standard treatments one defines the sets of components of the
electric and magnetic fields as
\begin{eqnarray}
E^{\mu } &=&F^{\mu \nu }t_{\nu },\quad B^{\mu }=(1/2)\varepsilon ^{\mu \nu
\lambda \sigma }F_{\lambda \sigma }t_{\nu }=(F^{\ast })^{\mu \nu }t_{\nu },
\nonumber \\
E^{0} &=&0,E^{i}=F^{i0};\quad B^{0}=0,B^{i}=(1/2)\varepsilon ^{0ikl}F_{lk},
\label{ebmi}
\end{eqnarray}
where $(F^{\ast })^{\alpha \beta }=(1/2)\varepsilon ^{\alpha \beta \gamma
\delta }F_{\gamma \delta }$ is the dual tensor (the components). (The
signature of the Minkowski tensor is $-2$, $\varepsilon ^{0123}=1,$ and $c=1$%
). \emph{The temporal components of }$E^{\mu }$\emph{\ and }$B^{\mu }$\emph{%
\ are zero. }Note that we can select a particular - but otherwise arbitrary
- inertial frame of reference (IFR) $S$ as the frame in which the relations (%
\ref{ebmi}) hold. $t^{\mu }$ can be interpreted as the 4-velocity (the
components in the ESC) of the observers that are at rest in $S$. In the
standard treatments the 3-vectors $\mathbf{E}$ and $\mathbf{B}$, as \emph{%
geometric quantities in the 3D space}, are constructed from the spatial
components $E^{i}$ and $B^{i}$ from (\ref{ebmi}) and \emph{the unit 3-vectors%
} $\mathbf{i},$ $\mathbf{j},$ $\mathbf{k,}$ e.g., $\mathbf{E=}F^{10}\mathbf{i%
}+F^{20}\mathbf{j}+F^{30}\mathbf{k.}$ These results are quoted in numerous
textbooks and papers treating relativistic electrodynamics in the TF, see,
e.g., [2,3]. Actually in the usual covariant approaches, e.g., [2,3], one
forgets about $E^{0}$ and $B^{0} $ components and simply makes the
identification of six independent components of $F^{\mu \nu }$ with three
components $E^{i}$, $E^{i}=F^{i0}$, and three components $B^{i}$, $%
B^{i}=(1/2)\varepsilon ^{ikl}F_{lk}.$ Since in SR we work with the 4D
spacetime the mapping between the components of $F^{\mu \nu }$ and the
components of the 3D vectors $\mathbf{E}$ and $\mathbf{B}$ is mathematically
better founded by the relations (\ref{ebmi}) than by their simple
identification. Therefore we proceed the consideration using (\ref{ebmi}).
Note that the whole procedure is made in an IFR with the ESC. In another
system of coordinates that is different than the ESC, e.g., differing in the
chosen synchronization (as it is the 'r' synchronization considered in [4]),
the identification of $E^{i}$ with $F^{i0},$ as in (\ref{ebmi}) (and also
for $B^{i}$), is impossible and meaningless. Further the components $E^{i}$
and $B^{i}$ are determined in the 4D spacetime in the standard basis $%
\left\{ \gamma _{\mu }\right\} .$ Thence when forming the geometric
quantities the components would need to be multiplied with the unit
4-vectors $\gamma _{i}$ and not with the unit 3-vectors.

Let us now apply the LT to the components given in (\ref{ebmi}). Under the
passive LT the sets of components $E^{\mu }$ and $B^{\mu }$ from (\ref{ebmi}%
) transform to $E^{\prime \mu }$ and $B^{\prime \mu }$ in the relatively
moving IFR $S^{\prime }$
\begin{eqnarray}
E^{\prime \mu } &=&F^{\prime \mu \nu }v_{\nu }^{\prime },\quad B^{\prime \mu
}=(1/2)\varepsilon ^{\mu \nu \lambda \sigma }F_{\lambda \sigma }^{\prime
}v_{\nu }^{\prime }=(F^{\ast })^{\prime \mu \nu }v_{\nu }^{\prime },
\nonumber \\
E^{\prime \mu } &=&\left( -\beta \gamma E^{1},\gamma
E^{1},E^{2},E^{3}\right) ,\ B^{\prime \mu }=\left( -\beta \gamma
B^{1},\gamma B^{1},B^{2},B^{3}\right) ,  \label{ebcr}
\end{eqnarray}
where $v_{\nu }^{\prime }=\left( \gamma ,\beta \gamma ,0,0\right) .$ The
unit 4-vector (the components) $t^{\mu }$ in the time direction in $S$
transforms upon the LT into the unit 4-vector $v^{\prime \nu }$, the
4-velocity of the moving observers, that contains not only the temporal
component but also $\neq 0$ spatial component. Thence, the LT transform the
set of components (\ref{ebmi}) into (\ref{ebcr}). Note that $E^{\prime \mu }$%
\emph{\ and }$B^{\prime \mu }$\emph{\ do have the temporal components as
well. }Further the components $E^{\mu }$ ($B^{\mu }$) in $S$ transform upon
the LT again to the components $E^{\prime \mu }$ ($B^{\prime \mu }$) in $%
S^{\prime }$; there is no mixing of components. Actually this is the way in
which every well-defined 4-vector (the components) transforms upon the LT. A
geometric quantity, an abstract tensor $E^{a},$ can be represented by CBGQs
in $S$ and $S^{\prime }$ (both with the ESC) as $E^{\mu }\gamma _{\mu }$ and
$E^{\prime \mu }\gamma _{\mu }^{\prime },$ where $E^{\mu }$ and $E^{\prime
\mu }$ are given by the relations (\ref{ebmi}) and (\ref{ebcr})
respectively. \emph{All the primed quantities (components and the basis) are
obtained from the corresponding unprimed quantities through the LT.} Of
course it must hold that
\begin{equation}
E^{a}=E^{\mu }\gamma _{\mu }=E^{\prime \mu }\gamma _{\mu }^{\prime },
\label{ea}
\end{equation}
since the components $E^{\mu }$ transform by the LT and the basis $\gamma
_{\mu }$ transforms by the inverse LT thus leaving the whole CBGQ\ invariant
upon the passive LT. The invariance of some 4D CBGQ upon the passive LT is
the crucial requirement that must be satisfied by any well-defined 4D
quantity. It reflects the fact that such mathematical, invariant, geometric
4D quantity represents \emph{the same physical object} for relatively moving
observers. The use of CBGQs enables us to have clearly and correctly defined
the concept of sameness of a physical system for different observers. The
importance of this concept in SR was first pointed out in [9,10]. However
they also worked with components in the ESC (the covariant quantities) and
not with geometric quantities (the invariant quantities). It should be noted
that in all other standard treatments, e.g., [1-3], the importance of such
concept is completely overlooked what caused many difficulties in
understanding SR. It can be easily checked by the direct inspection that (%
\ref{ea}) holds when $E^{\mu }$ and $E^{\prime \mu }$ are given by (\ref
{ebmi}) and (\ref{ebcr}). (The same holds for $B^{a}.$)

In contrast to the above consideration in all usual treatments, e.g., [2,3]
and [11] eqs. (3.5) and (3.24), in $S^{\prime }$ one again simply makes the
identification of six independent components of $F^{\prime \mu \nu }$ with
three components $E^{\prime i}$, $E^{\prime i}=F^{\prime i0}$, and three
components $B^{\prime i}$, $B^{\prime i}=(1/2)\varepsilon
^{ikl}F_{lk}^{\prime }.$ This means that standard treatments assume that
under the passive LT the set of components $t^{\mu }=\left( 1,0,0,0\right) $
from $S$ transform to $t^{\prime \nu }=\left( 1,0,0,0\right) $ ($t^{\prime
\nu }$ are the components of the unit 4-vector in the time direction in $%
S^{\prime }$ and in the ESC), and consequently that $E^{\mu }$ and $B^{\mu }$
from (\ref{ebmi}) transform to $E_{st.}^{\prime \mu }$ and $B_{st.}^{\prime
\mu }$ in $S^{\prime },$
\begin{eqnarray}
E_{st.}^{\prime \mu } &=&F^{\prime \mu \nu }t_{\nu }^{\prime
},B_{st.}^{\prime \mu }=(F^{\ast })^{\prime \mu \nu }t_{\nu }^{\prime };\
E_{st.}^{\prime \mu }=\left( 0,E^{1},\gamma E^{2}-\gamma \beta B^{3},\gamma
E^{3}+\gamma \beta B^{2}\right) ,  \nonumber \\
B_{st.}^{\prime \mu } &=&\left( 0,B^{1},\gamma B^{2}+\gamma \beta
E^{3},\gamma B^{3}-\gamma \beta E^{2}\right) ,  \label{kr}
\end{eqnarray}
where the subscript - st. is for - standard. \emph{The temporal components
of }$E_{st.}^{\prime \mu }$\emph{\ and }$B_{st.}^{\prime \mu }$\emph{\ in }$%
S^{\prime }$\emph{\ are again zero as are the temporal components of }$%
E^{\mu }$\emph{\ and }$B^{\mu }$\emph{\ in }$S.$\emph{\ }This fact clearly
shows that \emph{the transformations given by the relation} \emph{(\ref{kr})
are not the LT of some well-defined 4D quantities; the LT cannot transform a
4-vector for which the temporal component is zero in one frame }$S$\emph{\
to the 4-vector with the same property in relatively moving frame }$%
S^{\prime }$. Also \emph{the LT cannot transform the unit 4-vector in the
time direction in one frame} $S$ \emph{to the unit 4-vector in the time
direction in another relatively moving frame }$S^{\prime }.$ Obviously $%
E_{st.}^{\prime \mu }$ and $B_{st.}^{\prime \mu }$ are completely different
quantities than $E^{\prime \mu }$ and $B^{\prime \mu }$ (\ref{ebcr}) that
are obtained by the LT. We can easily check that
\begin{equation}
E_{st.}^{\prime \mu }\gamma _{\mu }^{\prime }\neq E^{\mu }\gamma _{\mu
},\quad B_{st.}^{\prime \mu }\gamma _{\mu }^{\prime }\neq B^{\mu }\gamma
_{\mu }.  \label{krive}
\end{equation}
This means that, e.g., $E^{\mu }\gamma _{\mu }$ and $E_{st.}^{\prime \mu
}\gamma _{\mu }^{\prime }$ \emph{are not the same quantity for observers in}
$S$ \emph{and} $S^{\prime }.$ As far as relativity is concerned the
quantities, e.g., $E^{\mu }\gamma _{\mu }$ and $E_{st.}^{\prime \mu }\gamma
_{\mu }^{\prime },$ are not related to one another. Their identification is
the typical case of mistaken identity. The fact that they are measured by
two observers does not mean that relativity has something to do with the
problem. The reason is that observers $S$ and $S^{\prime }$ are not looking
at the same physical object but at two different objects. \emph{Every
observer makes measurement on its own object and such measurements are not
related by the LT.} Thus the transformations (\ref{kr}) are not the LT and $%
E_{st.}^{\prime \mu }$ and $B_{st.}^{\prime \mu }$, in contrast to $%
E^{\prime \mu }$\ and $B^{\prime \mu },$\ are not well-defined 4D quantities.

From \emph{the relativistically incorrect transformations} (\ref{kr}) one
simply derives the transformations of the spatial components $%
E_{st.}^{\prime i}$ and $B_{st.}^{\prime i}$. As can be seen from (\ref{kr})
\emph{the transformations of} $E_{st.}^{\prime i}$ \emph{and} $%
B_{st.}^{\prime i}$ \emph{are exactly the standard transformations of
components of the 3-vectors} $\mathbf{E}$ \emph{and} $\mathbf{B}$ that are
obtained by Einstein in [1] and subsequently quoted in almost every textbook
and paper on relativistic electrodynamics. Then in the same way as in $S$
the 3-vectors $\mathbf{E}^{\prime }$ and $\mathbf{B}^{\prime }$ as \emph{%
geometric quantities in the 3D space} are constructed in $S^{\prime }$ from
the spatial components $E_{st.}^{\prime i}$ and $B_{st.}^{\prime i}$ and
\emph{the unit 3-vectors} $\mathbf{i}^{\prime },$ $\mathbf{j}^{\prime },$ $%
\mathbf{k}^{\prime }\mathbf{,}$ e.g., $\mathbf{E}^{\prime }\mathbf{=}%
F^{\prime 10}\mathbf{i}^{\prime }+F^{\prime 20}\mathbf{j}^{\prime
}+F^{\prime 30}\mathbf{k}^{\prime }\mathbf{.}$ Both the transformations (\ref
{kr}) and the transformations for $E_{st.}^{\prime i}$ and $B_{st.}^{\prime
i}$ (i.e., for $\mathbf{E}^{\prime }$ and $\mathbf{B}^{\prime }$) are
typical examples of the ''apparent'' transformations (AT) that are first
discussed in [9] and [10]. The AT of the spatial distances (the Lorentz
contraction) and the temporal distances (the dilatation of time) are
elaborated in detail in [4] and [8] (see also [12]), and in [4] I have
discussed the AT of $\mathbf{E}$ and $\mathbf{B}$. It is explicitly shown in
[8] that the true agreement with experiments that test SR exists only when
the theory deals with well-defined 4D quantities, i.e., the quantities that
are invariant upon the passive LT.

In all previous treatments of SR, e.g., [1-3] [11], the transformations for $%
E_{st.}^{\prime i}$ and $B_{st.}^{\prime i}$ are considered to be the LT of
the 3D electric and magnetic fields. However our analysis shows that the
transformations for $E_{st.}^{\prime i}$ and $B_{st.}^{\prime i}$ are
derived from \emph{the relativistically incorrect transformations} (\ref{kr}%
) and that the 3-vectors $\mathbf{E}^{\prime }$ and $\mathbf{B}^{\prime }$
are formed by an incorrect procedure in 4D spacetime, i.e., by multiplying
these \emph{relativistically incorrect components} with \emph{the unit
3-vectors.} All this together exactly proves that the standard
transformations for $\mathbf{E}^{\prime }$ and $\mathbf{B}^{\prime }$ have
absolutely nothing to do with the LT, and that the quantities $%
E_{st.}^{\prime i}$ and $B_{st.}^{\prime i},$ i.e., the 3-vectors $\mathbf{E}
$ and $\mathbf{B}$ are not well-defined 4D quantities. Consequently \emph{%
the usual ME with 3D }$\mathbf{E}$\emph{\ and }$\mathbf{B}$\emph{\ are not
in agreement with SR and they are not physically equivalent with
relativistically correct FE with }$F^{ab}$ (see also [4]).

The relation (\ref{ebmi}) reveals that we can always select a particular -
but otherwise arbitrary - IFR $S$ in which the temporal components of $%
E^{\mu }$ and $B^{\mu }$ are zero. Then in that frame the usual ME for the
spatial components $E^{i}$ and $B^{i}$ (of $E^{\mu }$ and $B^{\mu }$) will
be fulfilled. As a consequence the usual ME can explain all experiments that
are performed in one reference frame. However as shown above the temporal
components of $E^{\prime \mu }$ and $B^{\prime \mu }$ are not zero; (\ref
{ebcr}) is relativistically correct, but it is not the case with (\ref{kr}).
This means that the usual ME cannot be used for the explanation of any
experiment that test SR, i.e., in which relatively moving observers have to
compare their data \emph{obtained by measurements on the same physical
object.}

The relations (\ref{ebcr}) imply that in the ESC the well-defined 4D
electric and magnetic fields will be CBGQs, the 4-vectors, $E^{\mu }\gamma
_{\mu }=F^{\mu \nu }v_{\nu }\gamma _{\mu }$ and $B^{\mu }\gamma _{\mu
}=(F^{\ast })^{\mu \nu }v_{\nu }\gamma _{\mu }$ respectively. In an
arbitrary chosen IFR $S$ $v_{\nu }$ can be taken to be in the time
direction, i.e., $v_{\nu }=t_{\nu }$, whence one finds (\ref{ebmi}) in $S$
and (\ref{ebcr}) in any relatively moving IFR $S^{\prime }.$ (The components
in the ESC, e.g., $E^{\mu }=F^{\mu \nu }v_{\nu }$, and the covariant
formulation of electrodynamics with them, are considered in [13], [12] and
[14]). In order to have the electric and magnetic 4-vectors defined without
reference frames, i.e., \emph{independent of the chosen reference frame and
of the chosen system of coordinates in it}, we employ the abstract tensors
and write $E^{a}=F^{ab}v_{b}$ and $B^{a}=-(1/2)\varepsilon
^{abcd}v_{b}F_{cd},$ see [4,5] and [7]. The velocity $v_{b}$ and all other
quantities entering into these relations are defined without reference
frames. $v_{b}$ characterizes some general observer. Thus \emph{the
relations for} $E^{a}$ \emph{and }$B^{a}$ \emph{hold for any observer.} When
some reference frame is chosen with the ESC in it and when $v_{b}$ is
specified to be in the time direction in that frame, i.e., $v_{b}=t_{b}$,
then all results of the classical electromagnetism are recovered in that
frame. However, in contrast to the description of the electromagnetism with
the 3D $\mathbf{E}$ and $\mathbf{B,}$ the description with $E^{a}$ and $B^{a}
$ is correct not only in that frame but in all other relatively moving
frames and it holds for any permissible choice of coordinates. In [4] I have
also presented the form of the LT that is independent of the chosen
coordinates. Furthermore I have developed three equivalent but independent,
consistent and complete formulations of electrodynamics with abstract
tensors, with $F^{ab}$, $E^{a}$ and $B^{a},$ and with their complex
combination. (Note that the above relations for $E^{a}$ and $B^{a}$ are not
the physical definitions of $E^{a}$ and $B^{a}$ but they simply connect the
independent and complete formulations with $F^{ab}$ and $E^{a},$ $B^{a}$.
The physical definitions of $E^{a}$ and $B^{a}$ are given in terms of the
Lorentz force expressed with $E^{a}$ and $B^{a}$ and Newton's second law, as
in [4] in the TF and in the first paper in [6] in the CAF.)

The same situation as in TF happens in the standard CAF, e.g., [15,16]. The
ESC, i.e., the standard basis $\left\{ \gamma _{\mu }\right\} ,$ is chosen
from the outset and the relations for $E$ and $B$ are written explicitly
using the basis 1-vector in the time direction, $\gamma _{0}$. In some IFR$%
\;S,$ $E=F\cdot \gamma _{0}=F^{k0}\gamma _{k}$, and $B=-\gamma _{5}(F\wedge
\gamma _{0})=(1/2)\varepsilon ^{0ikl}F_{lk}\gamma _{i}$, where $\gamma _{5}$
is the pseudoscalar for the frame $\left\{ \gamma _{\mu }\right\} $, '$\cdot
$' and '$\wedge $' denote the inner and outer products of the basis
1-vectors. (This form for $E$ and $B$ is equivalent to the forms given in
the standard CAF, e.g., [15,16].) Then it is wrongly assumed that under the
active LT (expressed in CAF by rotors; see [15,16], [6]) the new, i.e., the
Lorentz transformed, $E^{\prime }$ and $B^{\prime }$ are $E_{st.}^{\prime
}=F^{\prime }\cdot \gamma _{0}$ and $B_{st.}^{\prime }=-\gamma
_{5}(F^{\prime }\wedge \gamma _{0}).$ However when $E$ is transformed by the
active LT then $R(F\cdot \gamma _{0})\widetilde{R}$ \emph{is not} equal to ($%
RF\widetilde{R})\cdot \gamma _{0}=E_{st.}^{\prime }$ (for the explicit form
of the rotor $R$ see, e.g., [16] ch.6 and [6]). Really the components of $%
E_{st.}^{\prime }$($B_{st.}^{\prime }$) are the same as in the AT (\ref{kr}%
), while the components of $RE\widetilde{R}$ ($RB\widetilde{R}$) are the
same as in the correct LT (\ref{ebcr}). Finally the standard transformations
for the 3D $\mathbf{E}$ and $\mathbf{B}$ are derived from the AT for $%
E_{st.}^{\prime }$ and $B_{st.}^{\prime }$, see [15] Sec. 18 and [16] Ch. 7.
In [6] I have presented the relativistically correct form for $E$ and $B$ by
replacing $\gamma _{0}$ with $v$, the velocity of some general observer,
which is also defined, as in TF, without reference frames. The CAF from [6]
enabled me to develop \emph{four} equivalent but independent, consistent and
complete formulations of electrodynamics with field bivector $F$, with
1-vectors $E$ and $B$ , with complex 1-vector $\Psi $ and, what is specific
for the CAF, with a \emph{real }Clifford multivector $\Psi $. Furthermore
all relevant quantities for the electromagnetism, the stress-energy 1-vector
$T(v)$, the energy density $U$ (scalar), the Poynting 1-vector $S$, the
angular momentum density $M$ (bivector) and the Lorentz force $K$ (1-vector)
are directly derived from the FE and all of them are defined without
reference frames.

The whole consideration explicitly shows that the 3D quantities $\mathbf{E}$
and $\mathbf{B}$, their transformations and the equations with them are
ill-defined in the 4D spacetime. More generally, the 3D quantities do not
have an independent physical reality in the 4D spacetime. Thence the
relativistically correct physics must be formulated with 4D quantities that
are defined without reference frames, or by the 4D CBGQs (e.g., as in [4,5],
[8] in the TF and [6] in the CAF). The principle of relativity is
automatically included in such theory with well-defined 4D quantities, while
in the standard approach to SR [1] it is postulated outside the mathematical
formulation of the theory. The comparison with experiments from [8] and [6]
reveals that the true agreement with experiments that test SR can be
achieved only when such well-defined 4D quantities are considered.\bigskip

\noindent \textbf{REFERENCES}\bigskip

\noindent $\left[ 1\right] $ A. Einstein, Ann. Physik. \textbf{17}, 891
(1905), tr. by W. Perrett and G.B.

Jeffery, in \textit{The Principle of Relativity} (Dover, New York).

\noindent $\left[ 2\right] $ J.D. Jackson, \textit{Classical Electrodynamics}
(Wiley, New York, 1977) 2nd edn..

\noindent $\left[ 3\right] $ L.D. Landau and E.M. Lifshitz, \textit{The
Classical Theory of Fields,} (Pergamon, Oxford,

1979) 4th edn..

\noindent $\left[ 4\right] $ T. Ivezi\'{c}, Found. Phys. \textbf{31}, 1139
(2001).

\noindent $\left[ 5\right] $ T. Ivezi\'{c}, Annales de la Fondation Louis de
Broglie \textbf{27}, 287 (2002).

\noindent $\left[ 6\right] $ T. Ivezi\'{c}, hep-th/0207250; hep-ph/0205277.

\noindent $\left[ 7\right] $ R.M. Wald, \textit{General Relativity} (The
University of Chicago Press, Chicago, 1984).

\noindent $\left[ 8\right] $ T. Ivezi\'{c}, Found. Phys. Lett. \textbf{15},
27 (2002); physics/0103026; physics/0101091.

\noindent $\left[ 9\right] $ F. Rohrlich, Nuovo Cimento B \textbf{45}, 76
(1966).

\noindent $\left[ 10\right] $ A. Gamba, Am. J. Phys. \textbf{35}, 83 (1967).

\noindent $\left[ 11\right] $ C.W. Misner, K.S.Thorne, and J.A. Wheeler,
\textit{Gravitation} (Freeman, San Francisco,

1970).

\noindent $\left[ 12\right] $ T. Ivezi\'{c}, Found. Phys. Lett. \textbf{12},
105 (1999); Found. Phys. Lett. \textbf{12}, 507 (1999).

\noindent $\left[ 13\right] $ H.N. N\'{u}\~{n}ez Y\'{e}pez, A.L. Salas
Brito, and C.A. Vargas, Revista Mexicana de F\'{i}sica

\textbf{34}, 636 (1988).

\noindent $\left[ 14\right] $ S. Esposito, Found. Phys. \textbf{28}, 231
(1998).

\noindent $\left[ 15\right] $ D. Hestenes, \textit{Space-Time Algebra }%
(Gordon and Breach, New York, 1966).

\noindent $\left[ 16\right] $ B. Jancewicz, \textit{Multivectors and
Clifford Algebra in Electrodynamics} (World

Scientific, Singapore, 1989).

\end{document}